\documentclass[12pt,preprint]{aastex}

\usepackage{psfig}
\newcommand{\Swift}{{\it Swift}}

\def\simlt{\mathrel{\hbox{\rlap{\hbox{\lower4pt\hbox{$\sim$}}}\hbox{$<$}}}}
\def\simgt{\mathrel{\hbox{\rlap{\hbox{\lower4pt\hbox{$\sim$}}}\hbox{$>$}}}}

\begin{document}

\shorttitle{Macronovae}
\shortauthors{ Kulkarni }

\title{Modeling  
Supernova-like Explosions Associated with Gamma-ray Bursts with Short Durations}

\author{\small
S. R. Kulkarni\altaffilmark{1}}
\altaffiltext{1}{Division of Astronomy, Mathematics \&\ Physics,
105-24, California Institute of Technology, Pasadena, CA 91125, USA}

\begin{abstract}

There is now good evidence linking short-hard GRBs with both
elliptical and spiral galaxies at relatively low redshifts, redshift
of about 0.2. This contrasts with the average redshift of about 2
of long-duration events, which also occur only in star-forming
galaxies.  The diversity of hosts is reminiscent of type Ia supernovae,
which are widely (but not universally) believed to originate from
the coalescence of white dwarfs. By analogy, it has been postulated
that short-hard bursts originate from neutron star mergers. Mergers,
as well as stellar core-collapse events (type II SNe and long-duration
GRBs) are accompanied by long-lived sub-relativistic components
powered by radioactive decay of unstable elements produced in the
explosion.  It is therefore interesting to explore whether short
duration events also have ejecta powered by radioactivity (i.e. that
are supernova-like).  Observations already inform us that any
supernova like component in the first few well studied short hard
bursts must be fainter than those typical of type Ia or core-collapse
supernovae.  Rather than refer to weaker  supernova-like component
as ``mini-super nova'', an etymologically indefensible term, I use
the term {\it macronova}.  I investigate the observability of
macronovae powered by neutron decay and by radioactive Nickel.
Separately, I note that a macronova will reprocess energetic emission
arising from a long lived central source.  I find that surprisingly
interesting limits on the basic parameters of macronovae can be
obtained provided observations are obtained with current 10-m class
telescopes  over a range of one hour to one day following the burst.
Detection or even strong constraints provided by macronova observations
can be expected to yield the most direct clue (barring data from
futuristic gravitational wave interferemeters) to the origin of
short hard bursts. It is this prize that, over the course of coming
few years, will undoubtedly motivate astronomers to mount ambitious
macronova campaigns.

\end{abstract}

\keywords{gamma rays:bursts ---  supernovae:general --- stars:novae}

\section{Background}
\label{sec:Background}

There is currently a great deal of excitement and sense of great
progress in our understanding of the mysterious class of short
duration, hard spectrum gamma-ray bursts.  GRB 050509b was the first
short hard burst to be localized to moderate precision ($<10''$).
This was made possible thanks to the prompt slewing of \Swift\ and
subsequent discovery of a rapidly fading X-ray source \citep{gsb+05}.
The X-ray afterglow was noted to coincide with a nearby cluster of
galaxies \citep{gsb+05} and lay in the outskirts of a bright
elliptical galaxy at $z=0.22$ \citep{bpp+05}.  Fortunately the next
two short hard bursts were localized to arcsecond accuracy thanks
to the discovery of their X-ray, optical or radio afterglows and,
as a result, greatly advanced the field of short hard bursts.
GRB\,050709 was detected by the {\it High Energy Transient Explorer}
\citep{vlr+05}.  The optical \citep{hwf+05} and X-ray \citep{ffp+05}
afterglow places this burst on a star-forming galaxy at $z=0.16$
\citep{ffp+05}.  GRB\,050724 was detected by {\it Swift} \citep{Covino05}.
The radio and optical afterglow localization places this event on
an elliptical galaxy \citep{bpc+05} at $z=0.26$ \citep{pbc+05}.

Collectively these three short bursts indicate both elliptical and
spiral host galaxies.  These associations lend credence to other
less well localized associations e.g.\ \citet{ngp+05}.  It is however
a recent development that is of potentially greater interest to
this paper, namely the claim by \cite{tcl+05} of a population of
short bursts that are even closer.

It is intriguing that the hosts of short hard bursts are as diverse
as those of Ia supernovae.  
A model (popular in the past but not ruled out by observations)
that is invoked for Ia
explosions involves double degenerate binary systems which at some
point coalesce and then explode. 
An early and enduring model for short hard bursts
is the coalescence of two neutron stars (or a neutron star and
a black hole). The coalescence is expected to produce a burst
of neutrinos, of gamma-ray and also eject neutron-rich ejecta
\citep{elp+89}.

Several years ago \citet{lp98} speculated the ejecta would 
result in a supernova-like explosion i.e.
a sub-relativistic explosion with radioactivity.  The purpose of
this paper is to revisit this topic now that the rough distance
scale has been determined.  Furthermore, the possibility that some
events may be even very close makes it doubly attractive to develop
quantitative models of associated supernova-like explosions.

Coalescence models are not silent on the issue of associated
non-relativistic outflows (e.g.\ \citealt{jr02,r05}). Indeed, there
appears to be a multiplicity of reasons for such outflows: tidal
tails (which appear inevitable in all numerical models investigated
to date), a wind driven by neutrino emission of a central massive
neutron star and explosion of a striped neutron star.  To this list
I add another possible mechanism: ejection of the outer regions of
the accretion disk owing to conservation of angular momentum.

The composition of the ejecta is a matter of fundamental theoretical
interest but also of great observational consequence.  For the
explosion to be observationally detected the ejecta must contain a
long-lived source of power.  Otherwise the ejecta cools rapidly and
is not detectable with the sensitivity of current telescopes.
Indeed, the same argument applies to ordinary SN.  Unfortunately,
theoretical models of coalescence offer no clear guidance on the
composition of the ejecta although the current prejudice is in favor
of neutron rich ejecta (see \citealt{frt99}).

The paper addresses two potentially novel ideas: an explosion in
which the decay of free neutrons provides the source of a long lived
source of power and the reprocessing of the luminosity of a long
lived central source into longer wavelength emission by the slow
moving ejecta.  A less speculative aspect of the paper is that I
take a critical look about photon production and photon-matter
equilibrium.  These two issues are not important for SN models (and
hence have not been addressed in the SN literature) but could be
of some importance for lower luminosity and smaller ejecta mass
explosions. Along the same vein, I have also investigated the
transparency of the ejecta to $\gamma$-rays -- an issue which is
critical given the expected low mass and high speed of the ejecta
(again relative to SN).

The two ideas discussed above (a neutron powered MN and long lived
central source) are clearly speculative. However, the model presented
here include the essential physics of such explosions but are
adequate to explore the feasibility of detecting associated explosions
over a wide range of conditions.

Now a word on terminology. It is clear from the observations of the
three bursts and their afterglows  that any accompanying sub-relativistic
explosion laced with or without radioactive isotopes is considerably
dimmer than typical supernovae (Ia or otherwise;
\citealt{bpp+05,hwf+05,ffp+05,bpc+05}) The word ``mini supernova''
may naturally come to one's mind  as an apt description of such low
luminosity explosions.  However, the juxtaposition of ``mini'' and
``super'' is not etymologically defensible, and will only burden
our field with more puzzling jargon.  After seeking alternative
names I settled on the word  {\it macronova}
(MN)\footnotemark\footnotetext{This word was suggested by P. A.
Price.} -- an explosion with energies between those of a nova and
a supernova and observationally distinguished by being brighter
than a typical nova ($M_V\sim -8\,$mag) but fainter than a typical
supernova ($M_V\sim -19\,$mag).

\section{The Physical Model}

All short hard burst models must be able to account for the
burst of gamma-ray emission. This requires ultra-relativistic
ejecta \citep{goodman86,paczynski86}. 
Here, we are focussed
entirely on possible ejecta but at sub-relativistic velocities.
The fundamental parameters of any associated macronova is the
initial internal energy (heat),
$E_0$, the mass
($M_{\rm ej}$) and the composition of the sub-relativistic 
ejecta.  Given that the
progenitors of short hard bursts are expected to be compact the
precise value of the initial radius, $R_0$, should not matter to
the relatively late time (tens of minutes to days) epochs of interest
to this paper.  Accordingly, rather arbitrarily, $R_0$ has been set
to $10^7\,$cm.  There is little guidance on $E_0$ and $M_{\rm ej}$
but $M_{\rm ej}\sim 10^{-4}\,M_\odot$ to $10^{-2}\,M_\odot$ (and
perhaps even $0.1\,M_\odot$) have been indicated by numerical studies
\citep{jr02,r05}.  Based on analogy with long duration GRBs, 
a reasonable value for $E_0$ is the isotropic
$\gamma$-ray energy release of the burst itself.  We set the fiducial
value for $E_0$ to be $10^{49}\,$erg.

It appears to me that there are three interesting choices for the
composition of the ejecta: a neutron rich ejecta in which the
elements decay rapidly (seconds; \citealt{frt99}), an ejecta dominated
by free neutrons and an ejecta dominated by $^{56}$Ni.  Isotopes
which decay too rapidly (e.g.\  neutron rich ejecta) or those which
decay on timescales much longer than a few days will not significantly
make increase the brightness of the resulting macronova.  A neutron-
and an $^{56}$Ni-MN are interesting in that if such events exist
then they are well suited to the timescales that are within reach
of current observations, 25th magnitude\footnotemark\footnotetext{
The following note may be of help to more theoretically oriented
readers. The model light curves presented here are for the Johnson
$I$ band (mean wavelength of $0.8\,\mu$) $I=0$ corresponds to
2550\,Jy. The AB system, an alternative system, is quite popular.
This system is defined by $m({\rm AB}) = -2.5\log10(f_\nu)-48.6$
where $f_\nu$ is the flux in the usual CGS units, erg cm$^{-2}$
s$^{-1}$. This corresponds to a zero point of 3630\,Jy at all
frequencies.  Thus 25th magnitude is a few tenths of $\mu$Jy.} on
timescales of hours to days.

The explosion energy, $E_0$, is composed of heat (internal energy)
and kinetic energy of the ejecta. The heat further drives the
expansion and so over time the internal energy is converted to
additional kinetic energy.  We will make the following, admittedly
artificial, assumption: the initial heat is much larger than the
initial kinetic energy.  The final kinetic energy is thus $E_0= 1/2
M_{\rm ej} v_s^2$ where $v_s$ is the final velocity of ejecta.
Consistent with the simplicity of the model we will assume that the
expanding ejecta is homogeneous.  The treatment in this paper is
entirely Newtonian (unless stated otherwise) but it is more convenient
to use $\beta_s=v_s/c$ rather than $v_s$.

At any given instant, the total internal energy of the expanding
ejecta, $E$, is composed of a thermal term, $E_{\rm th}$ arising
from the random motion of  the electrons (density, $n_e$) and ions
(density, $n_i$) and the energy in photons, $E_{\rm ph}$:
\begin{equation} E/V = \frac{3}{2} n_i (Z+1) k T + aT^4.  \label{eq:E}
\end{equation} Here, $V=4\pi R^3/3$, $N_i=M_{\rm ej}/(A m_H)$,
$n_i=N_i/V$ and $n_e=Zn_i$.  For future reference, the total number
of particles is $N=N_i(Z+1)$.  Implicit in Equation~\ref{eq:E} is
the assumption that the electron, ion and photon temperatures are
the same. This issue is considered in some detail in \S\ref{sec:neutron}.

The store of heat has gains and losses described by
\begin{equation}
\dot E = \varepsilon(t)M_{\mathrm{ej}} - L(t) - 4\pi R(t)^2 P v(t)
\label{eq:dotE}
\end{equation}
where $L(t)$ is the luminosity radiated at the surface and
$\varepsilon(t)$ is heating rate per gram from any source of energy
(e.g.\ radioactivity or a long-lived central source).  $P$ is the
total pressure and is given by the sum of gas and photon pressure:
\begin{equation}
P = n_i(Z+1) k T + a T^4/3.
\label{eq:P}
\end{equation}

As explained earlier, the ejecta gain speed rapidly from expansion
(the $4\pi R^2 P v$ work term).  Thus, following the initial
acceleration phase, the radius can be expected to increase linearly
with time:
\begin{equation}
	R(t) = R_0 + \beta_s c t.
\label{eq:R}
\end{equation}
With this (reasonable) assumption of coasting we avoid solving the
momentum equation and thus set $v=v_s$ in Equation~\ref{eq:dotE}.

Next,  we resort to the so-called ``diffusion'' approximation (see
\citealt{a96}; \citealt{p00}, volume II, \S 4.8),
\begin{equation}
L = E_{\rm ph}/t_{\rm d},
\label{eq:L}
\end{equation}
where 
\begin{equation}
t_d=B\kappa M_{\rm ej}/cR
\label{eq:td}
\end{equation} 
is the timescale for a typical photon to diffuse from the center
to the surface. The pre-factor $B$ in Equation~\ref{eq:td} depends
on the geometry and, following Padmanabhan ({\it ibid}), we set
$B=0.07$.  $\kappa$ is the mass opacity.

The composition of the ejecta determines the heating function,
$\varepsilon(t)$, and the emission spectrum. For neutrons, the
spectrum is entirely due to hydrogen.  Initially the photosphere
is simply electrons and protons and the main source of opacity is
due to Thompson scattering by the electrons.  The mass opacity is
$\kappa=\sigma_T/m_H= 0.4\,$cm$^{2}$\,g$^{-1}$; here $m_H$ is the
mass of a hydrogen atom and $\sigma_T$ is the Thompson cross section.
Recombination of protons and electrons begins in earnest when the
temperature reaches 20,000\,K and is complete by 5,000\,K.  With
recombination, the opacity from electrons disappears. Based on
models of hydrogen rich SNe we assume that the critical temperature,
the temperature at which electron opacity disappears, is $T_{\rm
RC}=10^4\,$K.

In contrast, for Nickel (or any other $Z\gg 1$ element) ejecta, the
spectrum will be dominated by strong metal lines (like Ia supernovae)
with strong absorption blueward of 4000\,\AA. Next, $Z/A\sim 0.5$
and $\kappa=0.2$\,cm$^2$~g$^{-1}$.   In this case, based on models
for Ia SNe spectra \citep{a96}, I assume that the photosphere is
entirely dominated by electrons for $T>T_{\rm RC}=5\times 10^3\,$K.

The Thompson optical depth is given by
\begin{equation}
\tau_{\rm es} = \frac{Z}{A}
                  \frac{M_{\rm ej}/m_H}{4\pi R^3/3}
                   R\sigma_T 
		= 20(M_{\rm ej}/10^{-3}M_\odot)(Z/A)R_{14}^{-2};
\label{eq:taues}
\end{equation}
here, we use the notation $Q_x$ as short hand notation for a physical
quantity normalized to $10^x$ in CGS units.  Thus the ejecta, for
our fiducial value of $M_{\rm ej}$, remain optically thick until
the size reaches about $10^{14}\,$cm. However, as noted above,
electron scattering ceases for $T<T_{\rm RC}$ following which the
use of Equation~\ref{eq:taues} is erroneous.  The reader is reminded
of this limitation in later sections where the  model light curves
are presented.

With the great simplification made possible by Equations~\ref{eq:R}
and \ref{eq:L}, the RHS of Equation~\ref{eq:dotE} is now a function
of $E$ and $t$.  Thus we have an ordinary differential equation in
$E$.  The integration of Equation~\ref{eq:dotE} is considerably
simplified (and speeded up) by casting $P$ in terms of $E$ (this
avoids solving a quartic function for $T$ at every integrator step).
From Equation~\ref{eq:P} we find that photon pressure triumphs over
gas pressure when the product of energy and radius $ER > \chi =
5\times 10^{55}((Z+1)/A)^{4/3} (M_{\rm
ej}/10^{-2}\,M_\odot)^{4/3}$\,erg\,cm.  The formula
$P=(1+\chi/(\chi+ER))E/3V$ allows for a smooth transition from the
photon rich to photon poor regime.

Applying the MATLAB ordinary differential equation solver,
\texttt{ode15s}, to Equation~\ref{eq:dotE} I obtained $E$ on a
logarithmic grid of time, starting from $10\,$ms to $10^7\,$s.  With
the run of  $E$ (and $R$) determined, I solved for $T$ by providing
the minimum of the photon temperature, $(E/aV)^{1/4}$ and the gas
temperature, $2E/(3Nk_B)$ as the initial guess value for the routine
\texttt{fzero} as applied to Equation~\ref{eq:E}.  With $T$ and $R$
in hand $E_{\rm ph}$ is easily calculated and thence $L$.

The effective temperature of the surface emission was computed using
the Stefan-Boltzman formula, $L = 4\pi R^2\sigma {T_\mathrm{eff}}^4$.
The spectrum of the emitted radiation was assumed to be a black
body with $T_{\mathrm{eff}}$.  Again this is a simplification in
that Comptonization in the photosphere has been ignored.

\section{Pure Explosion}
\label{sec:pure}

In this section we consider a pure explosion i.e. no subsequent
heating, $\varepsilon(t)=0$.  If photon pressure dominates then
$P=1/3 (E/V)$  and an analytical formula for $L(t)$ can be obtained
(Padmanabhan, {\it op cit}; \citealt{a96})\nocite{p00}:
\begin{equation}
L(t) = L_0 \exp\bigg(-\frac{t_h t + t^2/2}{t_h t_d(0)}\bigg);
\label{eq:Ltphot}
\end{equation}
here, $t_h=R_0/v_s$ is the initial hydrodynamic scale, $t_d(0)=
B(\kappa M_{\rm ej}/cR_0)$ is the initial diffusion timescale and
$L_0=E_0/t_d(0)$.  However, for the range of physical parameters
discussed in this paper, gas pressure could dominate over photon
pressure. Bearing this in mind, I integrated Equation~\ref{eq:dotE}
but with $P=2/3 (E/V)$ and found an equally simple analytical
solution:
\begin{equation}
L(t) = \frac{L_0}{(t/t_h+1)}
\exp\bigg(-\frac{t_h t + t^2/2}{t_h t_d(0)}\bigg).
\label{eq:Ltgas}
\end{equation}

\begin{figure}[th]
\centerline{\psfig{file=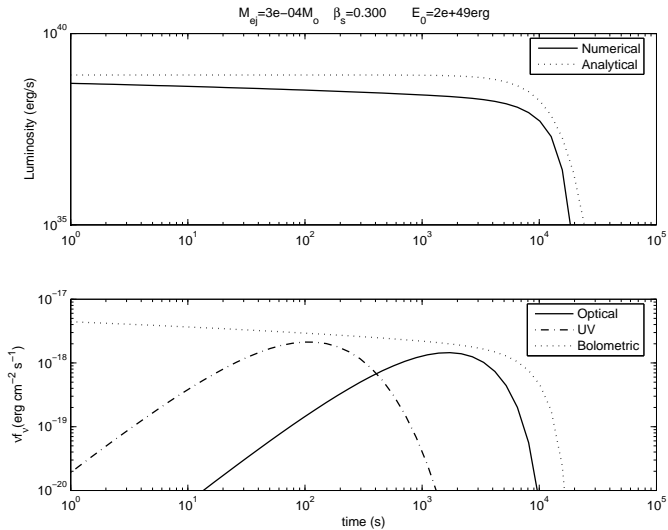,width=3.5in}}
\caption[]{\small
({\it Top}) The luminosity, $L(t)$, for an explosion with no heating
source. The model parameters ($M_{\rm ej}$ and $\beta_s$) are
indicated on the title line. The thick line is obtained from numerical
integration (Equation~\ref{eq:L}) whereas the dotted line is the
(photon dominated) analytical formula (Equation~\ref{eq:Ltphot}).
The numerical curve tracks the analytical curve (apart from a scaling
of 0.6); the two disagree as the MN evolves (see text).  ({\it
Bottom}) The optical and UV broad-band fluxes ($\nu f_\nu$) expected
for a macronova located at a redshift of $z=0.2$.  The dotted line
is the bolometric flux, $L(t)/(4\pi D^2)$ where $D=0.97\,$Gpc is
the distance. } 
\label{fig:Leak} 
\end{figure}

The analytical formula allow us to get some insight into the overall
behavior of the luminosity evolution.  First, we note that
$t_h=R_0/v_s=0.3\beta_s^{-1}\,$ms is much smaller than $t_d(0)=6.2\times
10^3 (M_{\rm ej}/10^{-3}\,M_\odot)\,$yr.  The internal energy, $E$,
decreases on the initial hydrodynamical scale and immediately
justifies our coasting approximation.  Next, the duration of the
signal is given by the geometric mean of $t_d(0)$ and $t_h$ and is
$\propto (M_{\rm ej}/v_s)^{1/2}$ but independent of $R_0$.  The
duration is not all that short, $\sim 0.3(M_{\rm
ej}/10^{-3}\,M_\odot)^{1/2}(\beta_s/0.1)^{-1/2}\,d$.  Third, the
peak emission, $E_0/t_d(0)= \beta_s^2 c^3 R_0/(2B\kappa)$, is
independent of the mass of the ejecta but directly proportional to
$R_0$ and the square of the final coasting speed, $v_s^2$.

Unfortunately, as can be seen from Figure~\ref{fig:Leak} (bottom
panel), model light curves\footnotemark\footnotetext{ The model
light curves presented here, unless stated otherwise, are for a
luminosity distance of 0.97 Gpc which corresponds to $z=0.2$ according
to the currently popular cosmology (flat Universe, Hubble's constant
of 71\,km\,s$^{-1}$\,Mpc$^{-1}$). The two wavebands discussed here
are the Optical (corresponding to restframe wavelength of
$8140\,\AA/(1+z)$) and the UV band (corresponding to restframe
wavelength of $1800\,\AA/(1+z)$). These two bands are chosen to
represent ground-based I-band (a fairly popular band amongst observers
given its lower immunity to lunar phase) and one of the UV bands
on the {\it Swift} UV-Optical telescope.  The time axis in all
figures has {\it not} been stretched by $1+z$.} for a macronova
located at $z=0.2$ is beyond current capabilities {\em in any band}
even in the best case (high $\beta_s$).  This pessimistic conclusion
is a direct result of small $R_0$, small $M_{\rm ej}$ and great
distance ($\sim 1\,$Gpc) for short hard bursts.

The situation is worse for lower shock speeds.  A lower shock speed
means a lower photon temperature and thus lower photon pressure.
The luminosity decreases by an additional factor  $\propto (t/t_h)^{-1}$
(Equation~\ref{eq:Ltgas}).  Next, the internal energy is shared
equitably between photons and particles and a lower photon density
means that a larger fraction of the internal energy is taken up the
particles.  Indeed, one can see a hint of the latter process in
Figure~\ref{fig:Leak} where the numerical curve (proceeding from
start of the burst to later times) is increasingly smaller than the
analytical curve. This is entirely a result of equipartition of $E$
between photons and particles and this equipartition is not accounted
for by in deriving Equation~\ref{eq:Ltphot}.

\section{Heating by Neutron Decay}
\label{sec:neutron}

On a timescale (half lifetime) of 10.4\,minutes, a free neutron
undergoes beta decay yielding a proton, a mildly relativistic
electron with mean energy, 0.3\,MeV) and an antineutrino.  Thus the
heating rate is entirely due to the newly minted electron (see
Appendix) and is
\begin{equation}
\varepsilon_n(t)= 3.2\times
10^{14}\,\mathrm{erg\,g^{-1}\,s^{-1}}.
\label{eq:varepsilon}
\end{equation}

Even though there are two unconstrained model physical parameters,
$M_{\rm ej}$ and $\beta_s$, a constraint may be reasonably imposed
by the prejudice that the macronova energy is, at best, comparable
to the gamma-ray energy release which we take to be $10^{49}\,$erg
\citep{ffp+05}.  Within this overall constraint I consider two
extreme cases for the neutron ejecta model: a high velocity case,
$\beta_s=0.3$ and a low(est) velocity case, $\beta_s=0.05$. The
mass of the ejecta was chosen so that $E_0\sim 10^{49}\,$erg.  The
range in $\beta$ is nicely bracketed by these two cases.  The escape
velocity of a neutron star is $0.3c$.  The energy released per
neutron decay of 0.3\,MeV  is equivalent to $\beta_s=0.025$.  Clearly,
final ejecta speeds below this value seem implausible (even such a
low value is implausible especially when considering that the ejecta
must first escape from the deep clutches of a compact object).  By
coincidence, as discussed below, these two cases nicely correspond
to a photon dominated and gas pressure dominated case, respectively.

\begin{figure}[thb]
\centerline{
\psfig{file=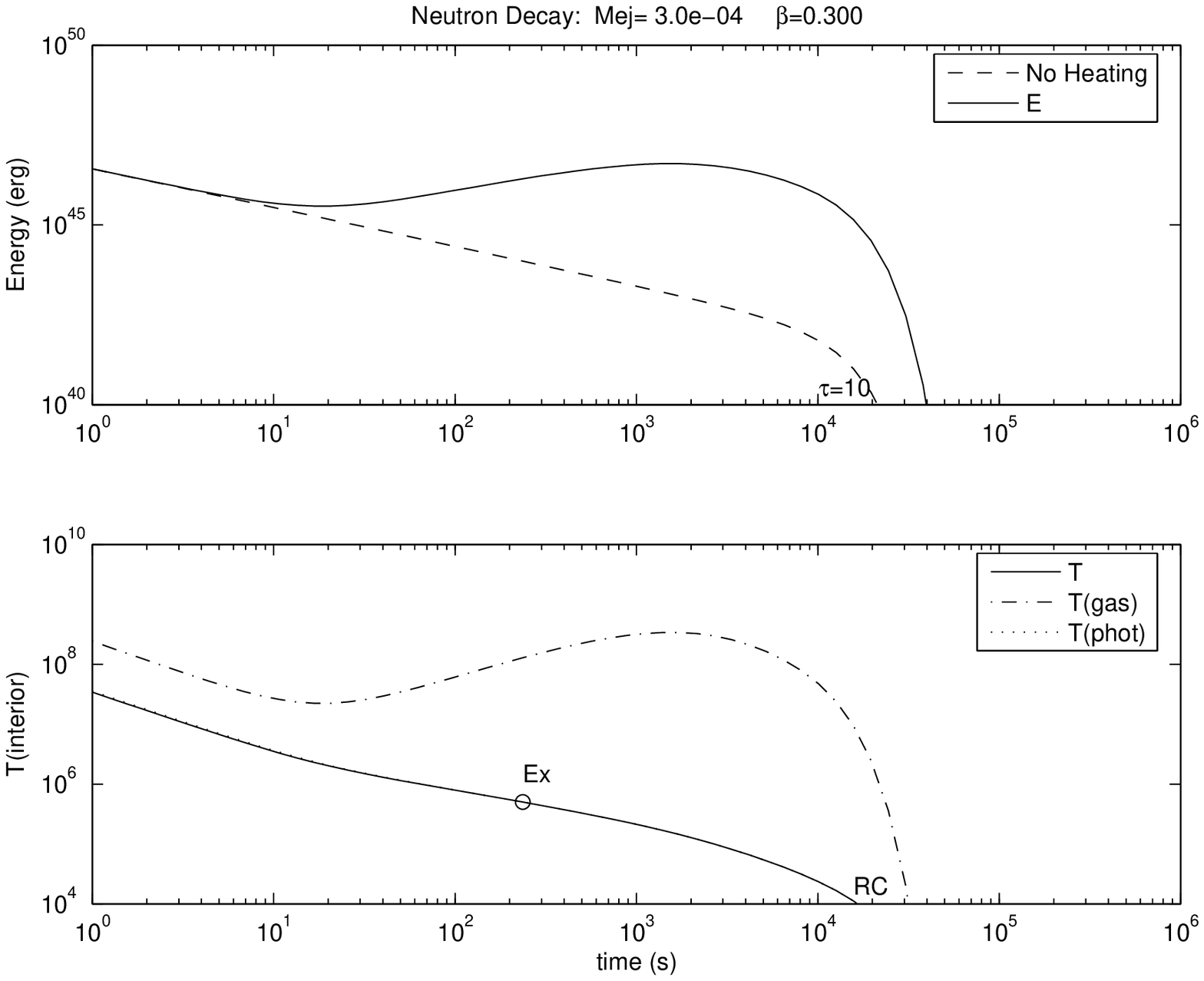,width=3in}\qquad
\psfig{file=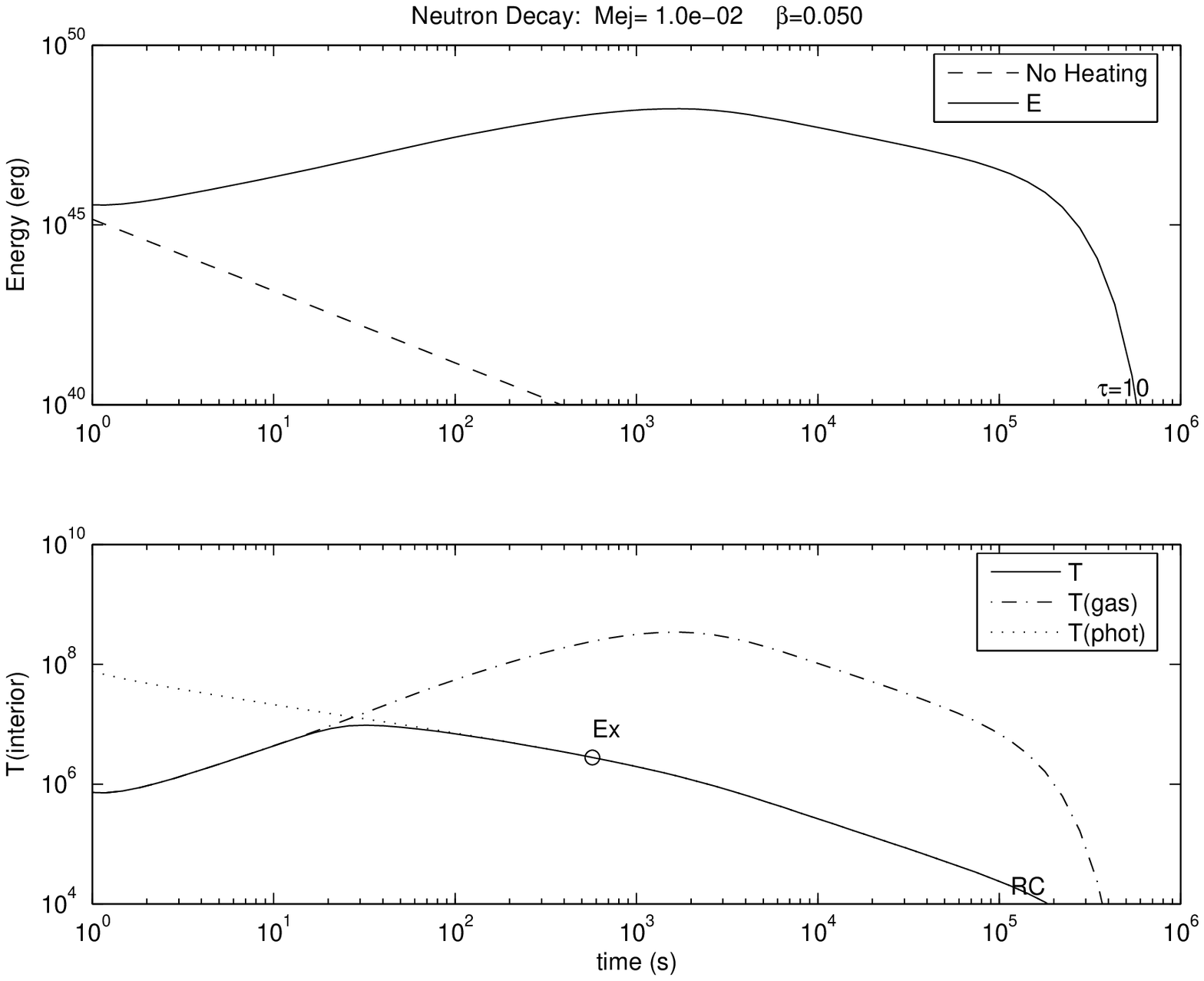,width=3in}}
\caption[]{\small 
({\it Top}) Internal energy ($E$) with (solid line) and without
(dashed line) neutron decay heating.  ({\it Bottom}) The interior
temperature obtained by solving Equation~\ref{eq:E} (solid line)
given $E$ and radius, $R$. The dash-dotted line is the temperature
one would obtain if all the internal energy was in particles whereas
the dotted line is that obtained if the photons dominated the
internal energy.  Each vertical pair of panels refers to a choice
of model parameters: coasting speeds of $\beta_s=0.3$ (left) and
$\beta_s=0.05$ (right). The explosion energy, $E_0=1/2v_s^2 M_{\rm
ej} \sim 2\times 10^{49}\,$erg in both cases.  ``RC'' marks the
epoch at which the surface temperature falls below $T_c=10^4\,$K
and $\tau=10$ marks the epoch at which electron scattering optical
depth is 10. ``Ex'' marks the epoch at which all the initial photons
are radiated away. }
\label{fig:NeutronHeating}
\end{figure}

The decay of neutrons extends the hot phase of the fireball expansion
(Figure~\ref{fig:NeutronHeating}).  This is nicely demonstrated in
Figure~\ref{fig:NeutronHeating} (right set of panels) where we see
that the decay of neutrons reverses the decrease in the internal
energy.  Indeed for the $\beta_s=0.05$ case neutron heating results
in the pressure switching from being dominated by gas to a  photon
dominated ejecta.  It is this gradual heating that makes a neutron
MN potentially detectable.

I now carry out a number of self-consistency checks.  To start with,
implicit in Equations~\ref{eq:E} and \ref{eq:P} is the assumption
that the ions, electrons and photons have the same temperature.
From the Appendix we see that the electron-electron and electron-ion
timescales are very short and we can safely assume that the electrons
and ions are in thermal equilibrium with respect to each other.

Next, I consider the time for the electrons to equilibrate with the
photons. There are two parts to this issue. First, the slowing down
of the ejected beta particle (electron).  Second, the equilibration
of thermal electrons with photons.  The energetic electron can be
slowed down by interaction with thermal electrons and also by
interaction with photons.  In the Appendix we show that the slowing
down and thermalization timescales have essentially the same value:
\begin{equation}
	t(\gamma,e)= \frac{3}{4}\frac{m_ec}{\sigma_T a T^4} =
        \Bigg(\frac{T}{2.7\times 10^5\,\mathrm{K}}\Bigg)^{-4}\,\mathrm{s}.
\label{eq:taugammae}
\end{equation}
Thus when the interior temperature falls below (say) $2.7\times
10^4\,$K the photon-electron equilibration time becomes significant,
about $10^4\,$s.  (However, by this time, most of the neutrons would
have decayed and all that realistically matters is the photon-matter
thermalization timescale).

An entirely independent concern is the availability of photons for
radiation at the surface. At the start of the explosion, by assumption,
all the internal energy, $E_0$, is in photons and thus the initial
number of photons is $N_{\mathrm{ph}}(0)=E_0/(2.7k_BT_0)$ where
$E_0=V_0 a T_0^4$, $V_0=4\pi/3 R_0^3$ is the initial volume and $a$
is the Stefan-Boltzmann radiation constant (see Appendix).  Photons
are continually lost at the surface.  Electron scattering does not
change the number of photons.  Thus an important self-consistency
check is whether the number of radiated photons is consistent with
the number of initial photons.  However, as can be seen from
Figure~\ref{fig:NeutronHeating} the number of radiated photons
exceeds the number of initial photons in less than ten minutes after
the explosion (the epoch is marked ``Ex'').  Relative to the
observations with large ground based telescopes this is a short
timescale and so it is imperative that we consider the generation
of new photons.

\subsection{Free-free emission as a source of new photons} For a
hot plasma with $A=Z=1$, the free-free process is the dominant
source of photon creation.  The free-free process is flat, $f_\nu
\propto \exp(-h\nu/k_BT)$ whereas the blackbody spectrum exhibits
the well known Planck spectrum. Thus the free-free process will
first start to populate the low-frequency end of the spectrum.  Once
the energy density, at a given frequency, reaches the energy density
of the black body spectrum at the same frequency, then free-free
absorption will suppress further production of free-free photons.

\begin{figure}[bht]
\centerline{
\psfig{file=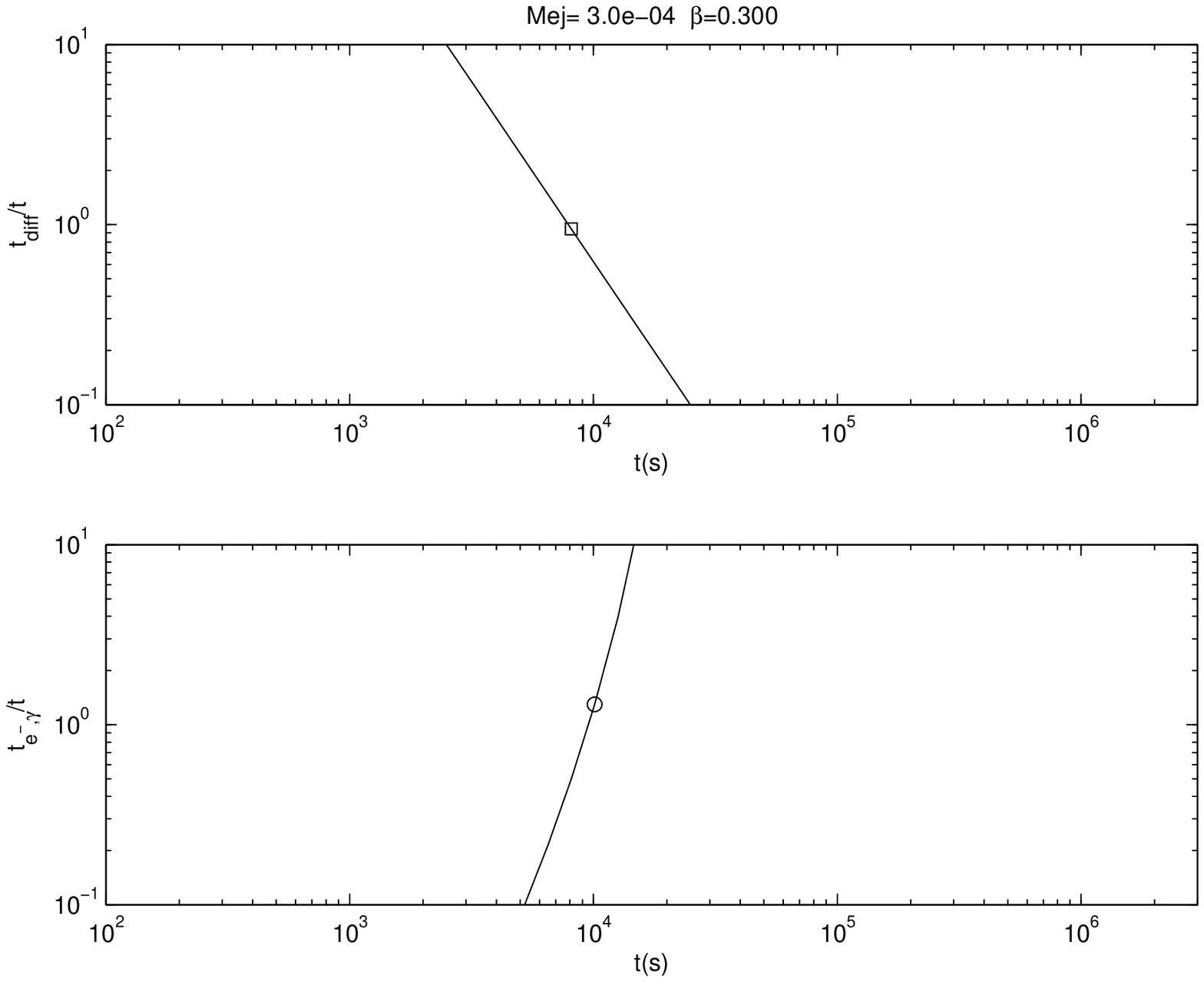,width=3in}\qquad
\psfig{file=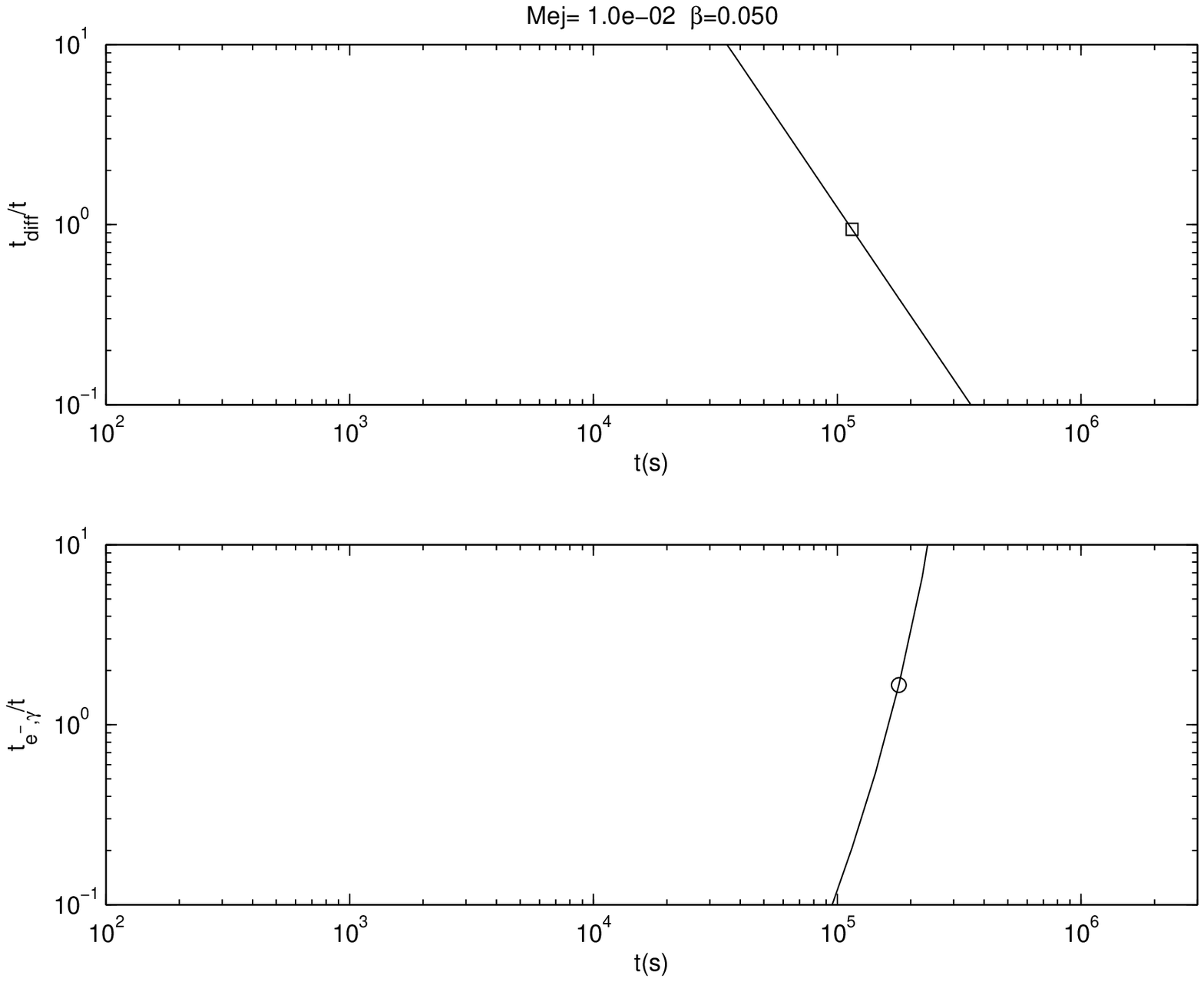,width=3in}}
\caption[]{\small 
Photon diffusion ($t_{\mathrm{d}}$) timescale and photon-electron
equilibration ($t_{e^,\mathrm\gamma}$) timescales, relative to the
expansion time ($t$), as a function of $t$.  The epoch at which
these timescales match the expansion time are marked by an open
square or open circle.  } 
\label{fig:timescales} 
\end{figure}

\begin{figure}[thb]
\centerline{
\psfig{file=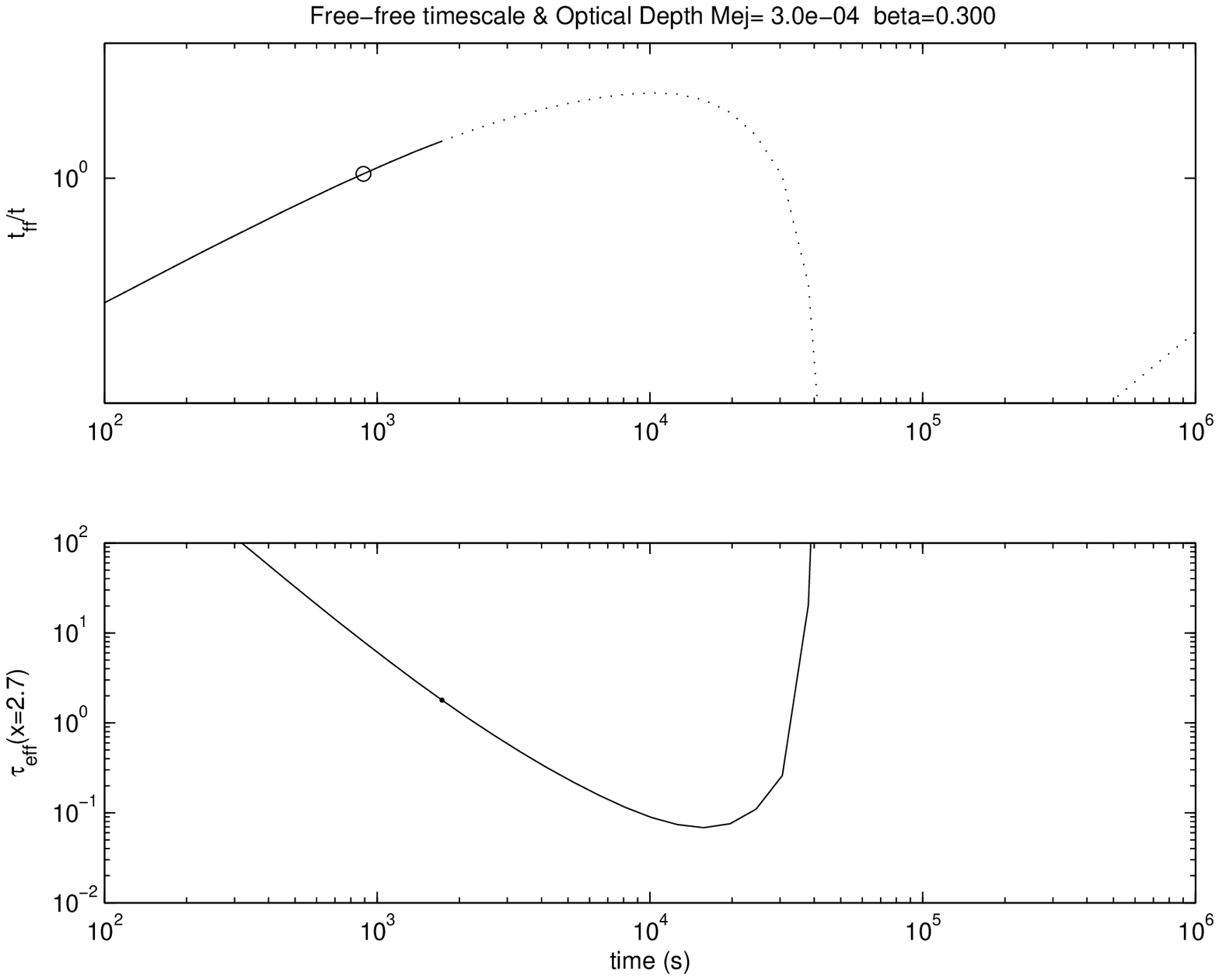,width=3in}\qquad
\psfig{file=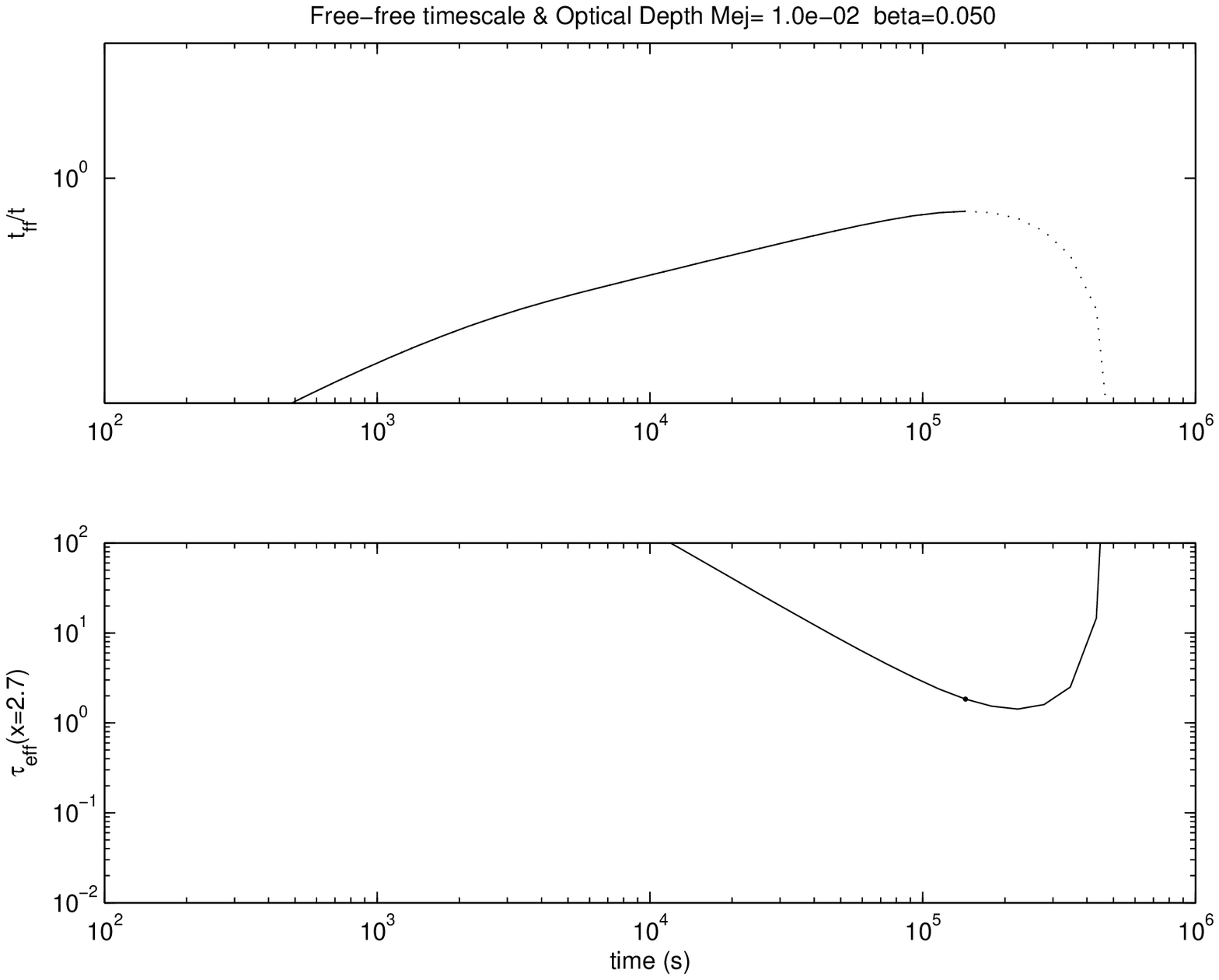,width=3in}}
\caption[]{\small
Run of the free-free timescale, $t_{\mathrm{ff}}$ ({\it top panel})
and the effective free-free optical depth ($\tau_*$) evaluated at
normalized frequency $x=h\nu/k_BT=2.7$ ({\it bottom panel}).  The
timescales are normalized by time past the explosion.  The free-free
timescale is only meaningful when the effective optical depth is
above unity (we have chosen $\tau_*=2$ as the criterion) and the
range of epochs for which this is the case is marked by a solid
line. The epoch at which free-free emission is unable to keep the
interior stocked with the blackbody photon density is marked by an
open square (top panel).  {\it Right.} The curves are for $\beta_s=0.05$
and $M_{\rm ej}=10^{-2}\,M_\odot$.  The macronova never gets optically
thin and hence the absence of squares.  {\it Left.} The curves are
for $\beta_s=0.3$ and $M_{\rm ej}=3\times 10^{-4}\,M_\odot$. Seemingly
$t_d$ rises at late times but this artifact is of little consequence
since there is no emission beyond $10^4\,$s.  } 
\label{fig:freefree}
\end{figure}

Provided the free-free optical depth, $\tau_{\mathrm{ff}}(\nu) =
R\alpha_{\mathrm{ff}} (\nu)$ (Appendix) is $\simgt 1$ for $\nu\simgt
kT/h$  a simple estimate for the timescale over which the free-free
process can build a photon density equal to that of the black body
radiation field (of the same temperature) is given by
\begin{equation}
    t_{\mathrm{ff}}=\frac{aT^4}{\epsilon_{\mathrm{ff}}(n_e,n_i,T)};
\label{eq:t_ff}
\end{equation}
here, $\epsilon_{\mathrm{ff}}(n_e,n_i,T)$ is the frequency integrated
free-free volume emissivity (see Appendix).  However, electron
scattering increases the effective optical depth of the free-free
process (\citealt{rl79}, p. 33)
\begin{equation}
    \tau_*(\nu) = \sqrt{\tau_{\mathrm{ff}}(\nu)(\tau_{\mathrm{ff}}(\nu)+\tau_{\mathrm{es}})}\ .
\label{eq:tau_*}
\end{equation}
$\tau_*$ takes into account that the relevant optical depth for any
emission process is the distance between the creation of a photon
and the absorption of the photon. Electron scattering increases the
probability of a newly minted photon to stay within the interior
and thereby increases the free-free optical depth.

Even after free-free process stops producing photons, the photons
in the interior are radiated on the photon diffusion timescale
(Equation~\ref{eq:td}). These two processes, the increase in the
effective optical depth and the photon diffusion timescale, prolong
the timescale of the macronova signal, making the macronova signal
detectable by the largest ground-based optical telescopes (which
unlike robotic telescopes respond on timescales of hours or worse).

In Figures~\ref{fig:timescales} and \ref{fig:freefree}, I present
various timescales ($t_{\rm d}$, $t_{\gamma,e}$, $t_{\mathrm{ff}}$)
and the frequency at which the effective free-free optical depth
(Equation~\ref{eq:tau_*}) is unity.  For $\beta_s=0.05$, $t_{\mathrm{ff}}$
is always smaller than $t$ and this means that the free-free process
keeps the interior well stocked with photons.  The remaining
timescales are longer, approaching a day.

For $\beta_s=0.3$ we find that $\tau_*$ as evaluated at $x=h\nu/k_BT=2.7$
falls below 2 at about 2,000 s.  Thus, for epochs smaller than
2,000\,s we can legitimately evaluate $t_{\mathrm{ff}}$. We find
that $t_{\mathrm{ff}}$ exceeds $t$ at about 1,000\,s (marked by
``ff''). We consider this epoch to mark the epoch at which free-free
emission ``freezes'' out in the sense that there is no significant
production of free-free emission beyond this epoch. However, the
photons in the interior leak out on the diffusion timescale which
becomes comparable to the expansion time at $t\sim 10^4\,$s.  In
summary, there is no shortage of photons for surface radiation for
ejecta velocity as high as $\beta_s=0.3$ but only for epochs earlier
than $10^4\,$s.

\begin{figure}[thb]
\centerline{\psfig{file=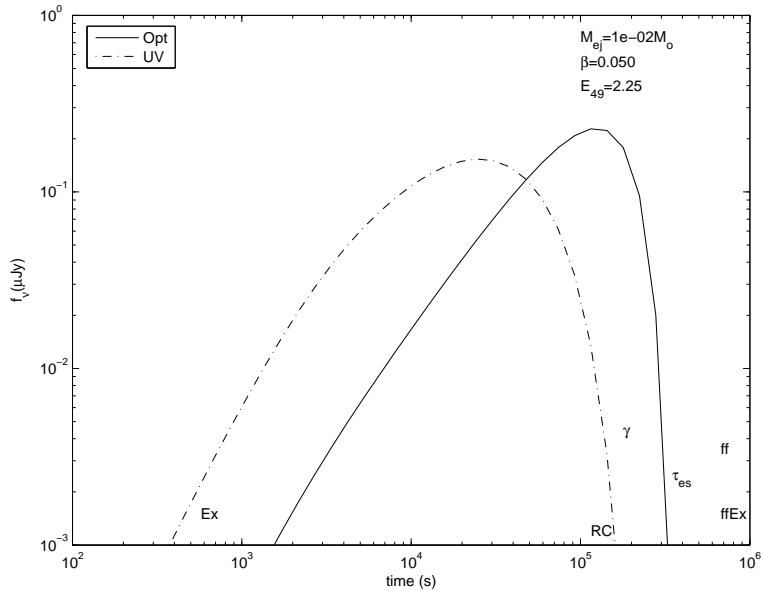,width=4in}}
\caption[]{\small
Optical and UV light curves for a macronova  located at a redshift,
$z=0.2$. The physical parameters, $M_{\rm ej}$ and $\beta$ are
displayed in the Figure.  The symbols at the bottom of the figure
are as follows: ``Ex'' (exhaustion of initial photons), ``ff''
(epoch beyond which free-free emission can no longer keep the photon
energy density at the black body energy density), ``$\gamma$''
(epoch at which electrons and photons decouple), ``$\tau_{\rm es}$''
(the electron scattering optical depth is 10 at this epoch), ``RC''
(the epoch at which the surface effective temperature is $10^4\,$K)
and ``ffEx'' is the epoch at which all the free-free photons generated
at the epoch marked ``ff'' are exhausted by radiation from the
surface.  In all cases the epoch is the left most character. Symbols
appearing close to either left or right vertical axis may have been
shifted to keep the symbols within the figure boundary.  
}
\label{fig:NeutronLCB} 
\end{figure}

\subsection{The Light Curves}

\begin{figure}[hbt]
\centerline{\psfig{file=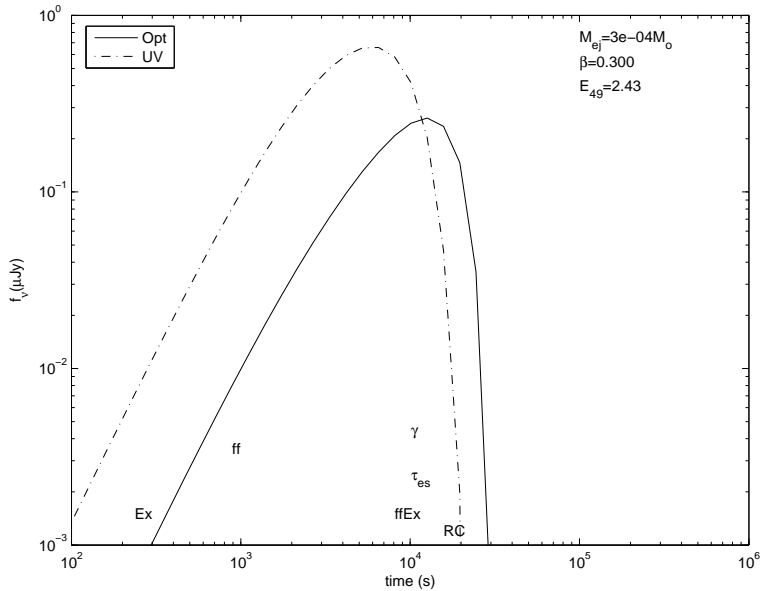,width=4in}}
\caption[]{\small
Expected light curve for a neutron decay powered macronova with
$\beta_s=0.3$. The explanations of the symbols can be found in the
caption to Figure~\ref{fig:NeutronLCB}.
}
\label{fig:NeutronLCA}
\end{figure}

The expected light curves for a macronova at $z=0.2$ are shown in
Figures~\ref{fig:NeutronLCB} and \ref{fig:NeutronLCA}.  For
$\beta=0.05$ the earliest constraining timescale is ``RC'' or the
epoch marking the recombination of the electrons with ions.  Beyond
day one, the model presented here should not be trusted.  For
$\beta=0.3$, as discussed earlier and graphically summarized in
Figure~\ref{fig:NeutronLCA}, the constraints come from photon
production. Complete photon exhaustion takes place at the start of
epoch of transparency (marked by ``$\tau_{\rm es}$'') and so emission
will cease rapidly at $10^4\,$s. Observations, if they have to any
value, must be obtained within an hour or two.

The peak flux in Figures~\ref{fig:NeutronLCB} and \ref{fig:NeutronLCA}
is about $0.3\,\mu$Jy.  With 15 minutes of integration time on a
10-m telescope one can easily detect this peak flux

Thus observations are capable of constraining a neutron powered MN
with explosion energy comparable to $10^{49}\,$erg (the typical
isotropic $\gamma$-ray energy release for short hard bursts).
However, these observations have to be obtained on timescales of
hours ($\beta_s=0.3$) to about a day ($\beta_s=0.05$).

Observations become less constraining with ejecta speeds larger than
that considered here because the macronova becomes transparent
earlier ($t\propto \beta_s^{-4}$, assuming that the explosion energy
is constant) and as a result photons are lost on a timescale approaching
the light crossing timescale i.e. rapid loss.

A potentially significant and non-trivial complication is confusion
of the macronova signal by the afterglow emission.  Optical afterglow
emission has been seen for GRB 050709 \citep{hwf+05}  and GRB 050724
\citep{bpc+05}.  Afterglow emission could, in principle, be
disentangled from the macronova signal by obtaining multi-band data.
Of particular value are simultaneous observations in the X-ray band.
No X-ray emission is expected in the macronova model.  Thus, X-ray
emission is an excellent tracer of the afterglow emission.  An
alternative source for X-rays is some sort of a central long-lived
source. As discussed in \S\ref{sec:reprocessor} a macronova would
reprocess the X-ray emission to lower energies. Thus, the X-ray
emission, at least in the macronova model, is a unique tracer of
genuine afterglow emission and as a  result can be used to distinguish
a genuine macronova signal from ordinary afterglow emission.

\section{Heating by Nickel}
\label{sec:nickel}

In the previous section I addressed in some detail photon generation
and photon-matter equilibration for $A=Z=1$ plasma.  Here, I present
light curves with Nickel ejecta.  The many transitions offered by
Nickel ($Z=28$, $A=56$) should make equilibration less of an issue.
Next, free-free mechanism becomes increasingly more productive for
higher $Z$ ions ($\propto Z^2$; see Appendix).  Thus the time for
free-free photons to be exhausted, a critical timescale should be
longer a Nickel MN (for a given $M_{\rm ej}$ and $\beta$) relative
to a neutron MN.

There is one matter that is important for a Nickel MN and that is
the issue of heating of matter. In a Nickel MN matter is heated by
deposition of gamma-rays released when Nickel decays to Cobalt.
Relative to the SN case, the energy deposition issue is important
for an MN both because of a smaller mass of ejecta and also higher
expansion speeds.

The details of Nickel $\rightarrow$ Cobalt $\rightarrow$ Iron decay
chain and the heating function are summarized in the Appendix.  In
view of the fact that most of our constraints come from early time
observations (less than 10 days) we will ignore heating from the
decay of Cobalt. Thus
\begin{equation}
\varepsilon_{\rm Ni}(t) = 
 3.9\times 10^{10}f_{\rm Ni}\exp(-\lambda_{\rm Ni}t)\,{\rm erg\,g^{-1} s^{-1}},
\end{equation}
where $\lambda_{\rm Ni}^{-1}=8.8\,$d and $f_{\rm Ni}$ is the mass
fraction of radioactive Nickel in the ejecta and is set to 1/3.

Nickel decay results in production of $\gamma$-rays with energies
between 0.158 and 1.56\,MeV (see Appendix).  \citet{cpk80} find
that the $\gamma$-ray mass absorption opacity is 0.029\,cm$^{2}$\,g$^{-1}$
for either the Ni$^{56}$ or Co$^{56}$ $\gamma$-ray decay spectrum.
Extraction of energy from the gamma-rays requires many scatterings
(especially for the sub-MeV gamma-rays) and this ``deposition''
function was first computed by \citet{cpk80}. However, the
$\varepsilon(t)$ we need in Equation~\ref{eq:dotE} is the deposition
function averaged over the entire mass of the ejecta. To this end,
using the local deposition function of \citet{cpk80}, I calculated
the bulk deposition fraction, $\eta_{\mathrm{es}}$ (for a uniform
density sphere) and expressed as a function of $\tau_{\rm es}$, the
center-to-surface Thompson optical depth.  The net effective heating
rate is thus $\eta_{\mathrm{es}}\varepsilon_{\rm Ni}(t)$.  I find
that only 20\% of the $\gamma$-ray energy is thermalized for
$\tau_{\rm es}=10$. Even for $\tau_{es}=100$ the net effective rate
is only 70\% (because, by construction, the ejecta is assumed to
be a homogeneous sphere; most of the $\gamma$-rays that are emitted
in the outer layers escape to the surface).

\begin{figure}[thb]
\centerline{\psfig{file=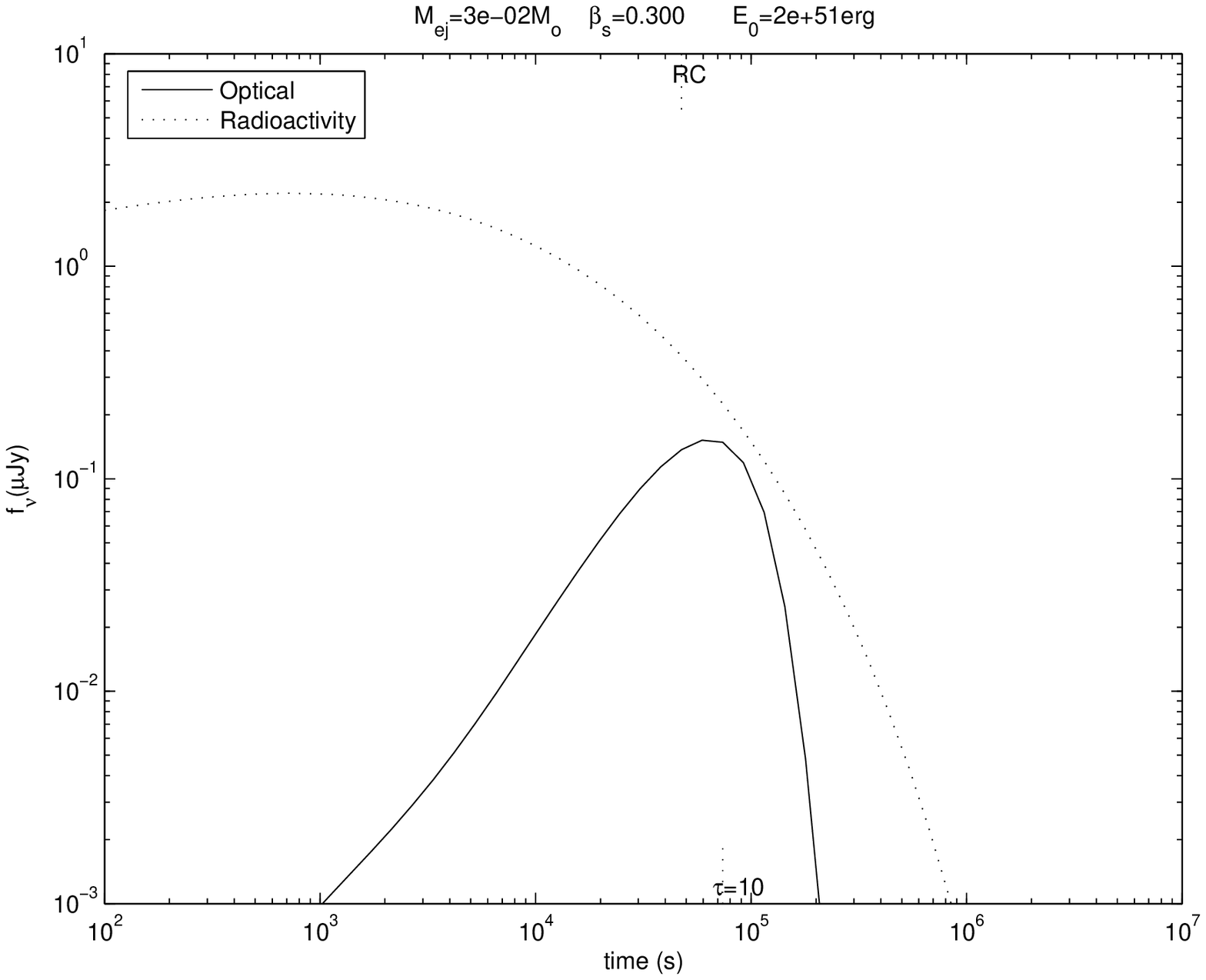,width=3in}
             \psfig{file=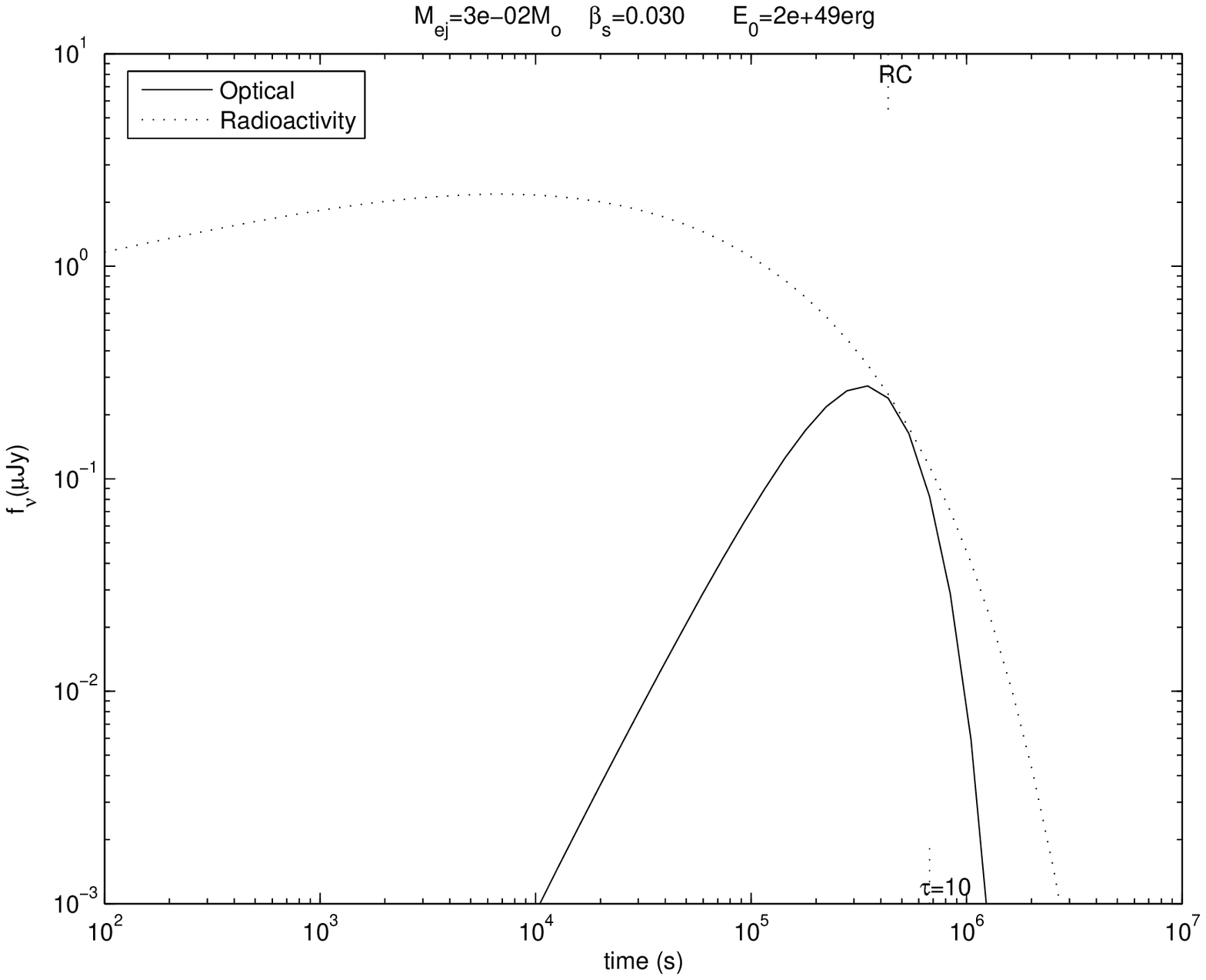,width=3in}}
\caption[]{\small
Model light curve for a
macronova at $z=0.22$ and powered by radioactive Nickel (one third
by mass).  The dotted line is the expected light curve if Nickel
radioactive decay power is instantly converted to the optical band.
The input includes only the fraction of $\gamma$-rays that are
absorbed and thermalized within the ejecta (i.e. $\varepsilon_{\rm
Ni}(t)\eta_{\rm es}$).  The slight curvature in the dotted line
(between $10^2\,$s and $10^4\,$s) is an artifact of the least squares
fit to the net energy deposition function (Equation~\ref{eq:eta}).
``RC'' is the epoch at which the effective surface temperature is
5,000\,K and the epoch at which the electron optical depth is 10
is marked. } 
\label{fig:Nickel} 
\end{figure}

The resulting lightcurves are plotted in Figure~\ref{fig:Nickel}
for $\beta_s=0.03$ and $\beta_s=0.3$.  I do not compute the UV light
curve given the high opacity of metals to the UV.  As with Ia
supernovae, peak optical emission is achieved when radioactive power
is fully converted to surface radiation \citep{a79,c81}.

$\beta_s=0.03$ is a sensible lower limit for the ejecta speed;
equating the energy released by radioactive decay to kinetic energy
yields $\beta_s=0.01$.  Peak optical emission is achieved at day
5. At this epoch, the photospheric temperature is in the vicinity
of $T_{\rm RC}=5\times 10^3\,$K. Thus the calculation should be
reasonably correct up to this epoch. Extrapolating from the neutron
MN case for similar model parameters I find that photon production
and equilibration are not significant issues.

For $\beta_s=0.3$, peak emission occurs at less than a day, again
closely coincident with the epoch when the effective photospheric
temperature is equal to $T_{\rm RC}$. For the neutron MN the
exhaustion of the free-free photons was the limiting timescale
($\sim 10^4\,$s). Qualitatively this timescale is expected to scale
as $Z$ (\S\ref{sec:nickel}) and if so the model light curve up to
the peak value is reliable.

To conclude, a Nickel powered MN is detectable only if the explosion
speed  is unreasonably low, $\beta_s=0.03$. Observations will not
place significant constraints for $\beta_s=0.3$. For such rapid
expansion, the MN suffers from lower deposition efficiency and an
onset of transparency before the bulk of Cobalt has decayed.
Conversely, there exists an opportunity for (futuristic) hard X-ray
missions to directly detect the gamma-ray lines following a short
hard burst!

\section{The MN as a reprocessor}
\label{sec:reprocessor}

The simplest idea for short hard bursts is a catastrophic explosion
with an engine that lives for a fraction of a second.  However, we
need to keep an open mind about the possibility that the event may
not necessarily be catastrophic (e.g.\ formation of a millisecond
magnetar) or that the engine or the accretion disk may live for a
time scale well beyond a fraction of a second.

It appears that the X-ray data  for GRB\,050709 are already suggestive
of a long lived source. A strong flare flare  lasting 0.1\,d and
radiating $10^{45}\,$erg energy (and argued not to arise in the
afterglow and hence by elimination to arise from a central source)
is seen sixteen {\it days} after the event \citep{ffp+05}. The
existence of such a flare in the X-ray band limits the Thompson
scattering depth to be less than unity at this epoch.  As can be
seen from Figures~\ref{fig:NeutronLCB} and \ref{fig:NeutronLCA}
this argument provides a (not-so interesting) lower limit to the
ejecta speed.

The main value of a macronova comes from the reprocessing of any
emission from a long-lived central source into longer wavelengths.
In effect, late time optical observations can potentially constrain
the heating term in Equation~\ref{eq:dotE}, regardless of whether
the heating arises from radioactivity or a long-lived X-ray source.
In this case, $\varepsilon(t)M_{\mathrm{ej}}$ refers to the luminosity
of the central source. The optical band is the favored band for the
detection of such reprocessed emission (given the current sensitivity
of facilities).

\begin{figure}[bth]
\centerline{\psfig{file=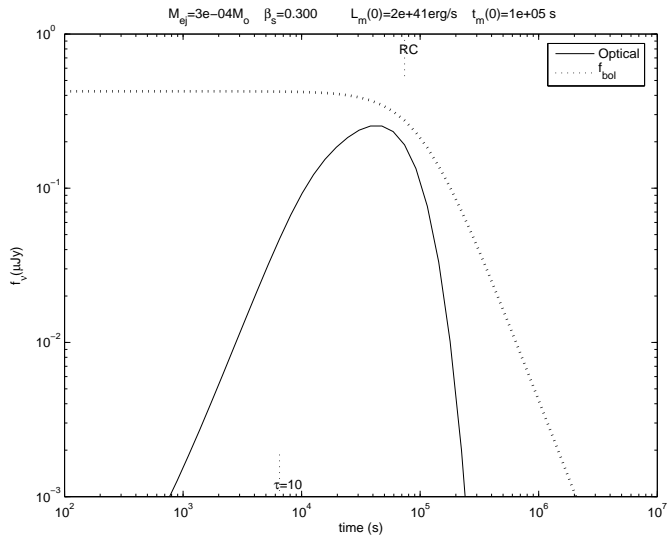,width=3.5in}}
\caption[]{\small
Expected optical light curve from a power law central source with
$L_m(t) = L_m(0)(1+t/t_m(0)^2)^{-1}$ with $L_m(0)=2\times 10^{41}\,$
erg s$^{-1}$ and $t_m(0)=10^5\,$s. The dotted line is the expected
flux if the power from the central source is instantly and fully
converted to the optical band. ``RC'' and ``$\tau=10$'' have the
same meanings as in previous figures.  
} 
\label{fig:Magnetar}
\end{figure}

A central magnetar is a specific example of a long lived central
source.  The spin down power of an orthogonal rotator is 
\begin{equation}
\varepsilon(t) = - \frac{B^2 R_n^6 \omega^4}{6c^3} 
\label{eq:magnetar}
\end{equation} 
where $B$ is the dipole field strength, $R_n$ is the radius of the
neutron, $\omega=2\pi/P$ is the rotation angular frequency and $P$
is the rotation period.  For $B=10^{15}\,$G, $R_n=16\,$km we obtain
$dE/dt \sim 10^{42} (P/100\,{\mathrm{ms}})^{-4}\,$erg s$^{-1}$ and
the characteristic age is $5\times 10^4\,$s. Constraining such a
beast (or something similar) is within the reach of current facilities
(Figure~\ref{fig:Magnetar}).

The power law decay discussed in this section is also applicable
for neutron rich ejecta. For such ejecta arguments have been advanced
that the resulting radioactive heating (from a variety of isotopes)
can be approximated as a power law (see \citet{lp98}).

Earlier we discussed of the potential confusion between a macronova
and afterglow emission. One clear distinction is that afterglow
emission does not suffer from an exponential cutoff whereas a
macronova dies down exponentially (when the optical depth decreases).
However, quality measurements are needed to distinguish power law
light curves of afterglow (flux $\propto t^\alpha$ with $\alpha=-1$
to $-2$) versus the late time exponential cutoff of macronova
emission.

\section{Conclusions}

The prevailing opinion is that short duration bursts arise from the
coalescence of a neutron star with another neutron star or a black
hole. The burst of $\gamma$-rays requires highly relativistic ejecta.
The issue that is very open is whether the bursts are also accompanied
by a sub-relativistic explosion.  This expectation is motivated
from numerical simulations of coalescence as well as the finding
of supernovae accompanying long duration bursts.  \citet{lp98} were
the first ones to consider the detection of optical signal from
sub-relativistic explosions.  Rather than referring to any such
accompanying nova-like explosion as a ``mini-supernova'' (a term
which only an astronomer will fail to recognize as an oxymoron) I
use the term macronova (and abbreviated as MN).

Essentially a macronova is similar to a supernova but with smaller
mass of ejecta. However, there are important differences. First,
supernova expand at relatively slow speed,  $\sim 10^9\,$cm\,s$^{-1}$
(thanks to the smalle escape velocity of a white dwarf).  Next,
radioactivity, specifically the decay of radioactive Nickel plays
a key role in powering supernovae.  The large mass of the ejecta,
the slow expansion and the 9-day and 111-day $e$-folding timescale
of Nickel and Cobalt (resulting  in a gradual release of radioactive
energy) all conspire to make supernovae attain significant brightness
on a timescale well suited to observations, namely weeks.

A macronova has none of these advantages. On general grounds, the
mass of ejecta is expected to be small, $10^{-2}\,M_\odot$. The
expansion speed is expected to be comparable to the escape velocity
of a neutron star, $0.3c$. Finally, there is no reason to believe
that the ejecta contains radioactive elements that decay on timescales
of days.

In this paper, I model the optical light curve for a macronova
powered by decaying neutrons or by decay of radioactive Nickel.
The smaller mass and the expected higher expansion velocities
necessitate careful attention to various timescales (photon generation,
photon-matter equilibrium, gamma-ray energy deposition).

Surprisingly a neutron powered  MN (with reasonable explosion
parameters) is within reach of current facilities. Disappointingly
a Nickel powered MN (with reasonably fast expansion speed) require
deeper observations to provide reasonable constraints. This result
is understandable in that a MN is not only a smaller supernova but
also a speeded up supernova.  Thus the detectability of a MN is
intimately connected with the decay time of radioactive elements
in the ejecta.  Too short a decay (seconds) or too long a decay
(weeks) will not result in a bright signal.  For example, artificially
changing the neutron decay time to 90\,s (whilst keeping the energy
released to be the same as neutron decay) significantly reduces the
signal from a macronova so as to (effectively) render it undetectable.
The difficulty of detecting Nickel powered MN illustrate the problem
with long decay timescales.

Next, I point out that a central source which lives beyond the
duration of the gamma-ray burst acts in much the same way as
radioactive heating.

Finally,  it is widely advertised that 
gravitational wave interferometers 
will provide the ultimate view of the collapse
which drive short hard bursts. However, these interferometers
(with appropriate sensitivity)
are in the distant future. Rapid response with large telescopes
has the potential to directly observe the debris of these
collapses -- and these observations can be done with currently
available facilities. This scientific prize is the
strongest motivation to mount ambitious campagins to detect
and study macronovae.

To date, GRB 050509b was observed rapidly (timescales of hours to
days) with the most sensitive facilities  (Keck, Subaru and HST).
The model developed here has been applied to these data and the
results reported elsewhere.

\acknowledgements

This paper is the first theoretical (even if mildly so) paper written
by the author.  The pedagogical tone of the paper mainly reflect
the attempts by the author (an observer) to make sure that he
understood the theoretical underpinnings.  I gratefully acknowledge
pedagogical and fruitful discussions with P. Goldreich, R.  Sari
and S. Sazonov.  I would like to thank L. Bildsten for acting as
the internal referee, U. Nakar for help in understanding the ODE
solver in MATLAB, R. Chevalier, T. Piran, A. MacFadyen, B. Schmidt
and R. Sunyaev for suggestions and clarifying discussions.  The
author is grateful for Biermann Lecture program of the Max Planck
Institute for Astrophysics, Garching for supporting a one month
sabbatical stay.  I am very grateful to T.-H Janka for patient
hearing and encouraging me to submit this paper.  The author
acknowledge financial support from a Space Telescope Science Institute
grant (HST-GO-10119) and the National Science Foundation.

\bibliographystyle{apj}
\bibliography{journals,refshb,macronova}

\section*{Appendix: Formulae \&\ Approximations used in the paper}

\subsection*{Blackbody Radiation}

The blackbody spectral intensity is 
\begin{equation}
B_\nu(T) = \frac{2}{\lambda^2}\frac{h\nu}{\exp({-h\nu/kT})-1}\,\,
\mathrm{erg\,cm}^{-2}\,\mathrm{s}^{-1}\,\mathrm{ster}^{-1}\,\mathrm{Hz}^{-1}.
\label{eq:B_nu}
\end{equation}
Frequently it helps to use physically normalized frequency unit,
$x=h\nu/(k_BT)$. Then,
\begin{equation}
B_x(T) = \frac{2h}{c^2} \Big( \frac{kT}{h}\Big)^4 \frac{x^3}{\exp(x)-1}.
\end{equation}
The spectral energy density is
$$
u_\nu(T) = \frac{4\pi}{c}B_\nu(T).
$$
The photon number density is 
\begin{eqnarray}
n(<\nu) &=& \int_0^\nu \frac{4\pi}{c}\frac{B_\nu(T)}{h\nu}d\nu \\
n(<x) &=& 8\pi\Big(\frac{kT}{h}\Big)^3\int_0^x \frac{x^2}{\exp(x)-1}dx \\
&=& 8\pi\Big(\frac{kT}{hc}\Big)^3\Gamma(3)\zeta_3(x)\\
&=&16\pi \Big(\frac{kT}{hc}\Big)^3\zeta_3(x);
\end{eqnarray}
where $\zeta_n(x)\equiv \Gamma(n)^{-1}\int_0^x t^{n-1}/(e^t-1)dt$
is an incomplete $\zeta$ function of order $n$. We note that
$\zeta_3(\infty)=1.202$ and $\zeta_4(\infty)=\pi^4/90$.  Thus the
number of photons per unit volume is $n_0= 19.2\pi (kT/hc)^3$. The
mean energy of the photon defined as $u(T)/n_0$ is $2.7kT$.  In the
Rayleigh-Jeans approximation, we obtain
\begin{equation}
n(<x) = 16\pi\Big(\frac{kT}{hc}\Big)^3\frac{x^2}{4} 
\end{equation}
This equation demonstrates that most of the photons have $x>1$ (with
less than 20\% of photons with $x<1$).  The fiftieth percentile is
$x=2.36$.

\subsection*{Free-free Emission}

The free-free volume emissivity rate (Rybicki \&\ Lightman 1979, p. 160) is 
\begin{equation}
\epsilon_{\mathrm{ff}}(\nu) = 6.8\times 10^{-38}Z^2 n_e n_i T^{-1/2}\exp(-h\nu/k_BT) 
\bar{g}_{\mathrm{ff}}\,\,\mathrm{erg}\,\mathrm{cm}^{-3}\mathrm{s}^{-1}\mathrm{Hz}^{-1}.
\label{eq:J_nu}
\end{equation}
Here, $\bar{g}_{\mathrm{ff}}$ is the velocity-averaged Gaunt factor
and varies from 5 to 1 as $u=h\nu/k_BT$ ranges from $10^{-4}$ to
1.  It appears to me that $\bar{g}_{\mathrm{ff}}=-\log(u)+1$ provides
a reasonable approximation (for $u<1$).  The frequency integrated
emissivity ($Z=A=1$)
\begin{equation}
\epsilon_{\mathrm{ff}} = 1.4\times 10^{-27}n_e^2 T^{1/2}\bar{g}_B
\,\,\mathrm{erg\, cm}^{-3}\,\mathrm{s}^{-1}
\label{eq:J}
\end{equation}
where $\bar{g}_B$ is the frequency-averaged Gaunt factor and a value
of 1.2 gives 20\% accuracy. For very hot plasma, relativistic effects
become important and the rate increases by a factor of $1 + T/2.2\times
10^9\,\mathrm{K}$.

The free-free density absorption coefficient, $\alpha_\nu(\mathrm{ff})$
$=n \sigma_\nu (\mathrm{ff})$ is given ({it op cit.}, p. 162)
\begin{equation}
\alpha_\nu(\mathrm{ff}) = 3.7\times 10^8 T^{-1/2} Z^2 n_e n_i \nu^{-3} 
(1-\exp(-h\nu/k_BT))\,\mathrm{cm}^{-1}
\bar{g}_{\mathrm{ff}}
\end{equation}

As noted in the previous section, it is useful to work in normalized
frequency units, $x=h\nu/kT$. Noting $\epsilon_{\mathrm{ff}}(x) =
(kT/h) \epsilon_{\mathrm{ff}}(\nu) $ we find ($Z=A=1$)
\begin{equation}
\epsilon_{\mathrm{ff}}(x)= 1.42\times 10^{-27} n^2T^{1/2} \exp(-x)
\,\mathrm{erg\, cm}^{-3}\,\mathrm{s}^{-1}.
\end{equation}

The number of photons above a value of $x$ is given by the integral
$y(x)=\int_x^\infty t^{-1}\exp(-t) dt$. Though simple looking, this
integral has no solution with known functions.  The usual (sensible)
approximation recognizes that there are few photons with $x>1$ and
thus the upper limit can be replaced by unity. The next approximation
is replace the exponential by unity. Thus $y(x)=-\log(x)$ (with
implicit understanding $x<1$). Alternatively, an empirical function
(obtained by numerical fitting),
\begin{equation} 
y(x) = 0.306 -0.413\log(x)+ 0.094 \log(x)^2,
\end{equation} 
provides a reasonable fit for $10^{-2}<x<5$.

\subsection*{Photon-electron equilibration timescale}

For a pure $A=Z=1$ gas, the primary primary photon-matter interaction
for very hot gas is through Thompson scattering of electrons.  Here
we summarize the basic physics of the timescale of thermalizing a
fast moving electron by the radiation field as well as the thermal
photon-thermal electron equilibration timescale.

Consider the fate of the electron ejected as a result of neutron
decay. The initial Lorentz factor of the electron ($\gamma_0$) can
range from 3.5 to unity with a mean value of 1.6.  The typical
ejected electron is much hotter than the radiation field (temperature
$T$) and will cool down by inverse scattering the ambient photons.
The rate of energy loss is
 \begin{equation}
       \frac{dE}{dt} = -\frac{4}{3}\sigma_T c \gamma^2\beta_s^2
       U_{\rm ph}.
  \label{eq:dEdtelectron} 
\end{equation}
where  the kinetic energy of the electron at any instant is given
by $E=(\gamma-1)m_e c^2$. Simplifying I obtain 
  \begin{equation}
     \frac{d\gamma}{dt} = A (1-\gamma^2) 
     \label{eq:dgammadt} 
   \end{equation}
where $A= {4\sigma_T c U_{\rm ph}}/{3m_e c^2}$.  Since $\int
dx(1-x^2)^{-1}=(1/2)\log((x+1)/(x-1))$ we find that the electron
loses a large fraction of energy over timescales of $A^{-1}$.

Next we consider an electron with energies comparable to that of
photons.  Consider an electron moving with velocity $v$ with respect
to the radiation field. In the electron rest frame the radiation
temperature is Doppler shifted as follows:
  \begin{equation}
    T(\theta) = T (1 + v/c) \cos(\theta)
  \end{equation}
and correspondingly the total intensity also has an angular dependence
given by
  \begin{equation}
    I(\theta) = \frac{ac}{4\pi} T(\theta)^4.
  \end{equation}
The momentum is $I(\theta)/c$ and the rate of momentum transfer to
the electron is
  \begin{eqnarray}
     F &=& - \int \sigma_T \frac{I(\theta)}{c} \cos(\theta)d\Omega \\
       &=& -\frac{4}{3}\sigma_T a T^4 \frac{v}{c}
  \end{eqnarray}

Averaging the velocity over the thermal distribution,
the work done by the radiation drag on the electron is
\begin{eqnarray}
\langle -Fv\rangle &=& \frac{4}{3}\sigma_T T^4 \langle\frac{v^2}{c}\rangle\\
                   &=& 4\sigma_T aT^4 \frac{k T_e}{m_e c}
\end{eqnarray}
where we have made use of the well known formula $1/2 m_e \langle
v^2\rangle = 3/2 k T_e$. However, the electrons must be getting
heated up by fluctuations in the radiation field because in
equilibrium, when $T=T_e$, the electron will neither gain energy
nor lose (on average). Thus the heat equation for the electron is
\begin{equation}
\frac{dQ}{dT} = 4\sigma_T a T^4 k (T-T_e)/m_e c.
\end{equation}
For fully ionized matter, $Q=3kT_e$ and thus we get the equation
well known to cosmologists \citep{p71}
\begin{equation}
	\frac{dT_e}{dt} = -A (T_e-T)
\end{equation}
where $U_{\rm ph}=aT^4$.  The thermalization timescale is thus
$\sim A^{-1}$.

\subsection*{Electron and Ion Equilibration}

A fast electron will lose energy to slow moving electrons by
Rutherford scattering.
From Padmanabhan (2000, volume I, pp. 439) we see
\begin{eqnarray}
t(e,e) &=& \frac{1}{4\pi}\frac{m_e^{1/2}(k_B T_e)^{3/2}}{e^4nL_e} 
          =  10^{-2}T_{e,7}^{3/2}n_{11}^{-1}\,\mathrm{s}, \\
t(i,i) &=& \frac{1}{4\pi}\frac{M^{1/2}(k_B T_i)^{3/2}}{e^4nL_i}
          = 4.3\times 10^{-1} T_{i,7}^{3/2} n_{11}^{-1}\,\mathrm{s},\ {\rm and} \\
t(e,i) &=& \Bigg(\frac{9\pi}{8}\Bigg)^{1/2}\frac{1}{Z^2}\Bigg(\frac{M}{m_e}\Bigg)t(e,e)
          = 10 Z^{-2}T_{e,7}^{3/2}n_{11}^{-1}\,\mathrm{s}.
\end{eqnarray}
Here $M$ is the mass of the ion and $L_e$ ($L_i$) refers to the
usual logarithmic factor for electrons (ions) and set to 10. $n$
is the particle density (cm$^{-3}$) and the density normalization
of $10^{11}\,$cm$^{-3}$ is appropriate for radius of $10^{14}\,$cm
and ejecta mass of $M_{\rm ej}=10^{-3}\,M_\odot$.

\subsection*{Heating by Neutron Decay}

A free neutron decays to a proton, electron and antineutrino.  The
half lifetime of the decay is 10.4 minutes and thus the $e$-folding
time, $t_N=900\,$s.  The mass difference between the neutron and
that of the electron and proton is $Q=0.782$\,MeV (see for example,
\citealt{k88}).

The sum total linear momentum is conserved and the energy carried
off by the electron and antineutrino must add up to $Q$. This would
nominally suggest that there are five remaining free parameters
($3\times 3$ momenta minus four constraints).  Thus the output of
decay can now be described by five free parameters and we choose
these to be the momentum of the electron, ${\mathbf p}$, and two
momenta of the antineutrino, ${\mathbf{q}}$.

The proton, by the virtue of its mass, obtains very little energy
in the process. Thue the energy of the antineutrino is $E_{\bar\nu}
= Q-T$ where, following the convention in nuclear physics, $T=E-m_ec^2$
is the kinetic energy and $E^2 = p^2 c^2 + m_e^2 c^4$ (here
$p=\sqrt{\mathbf{p} \cdot \mathbf{p}}$).  The magnitude of the
momentum of the antineutrino is $q=E_{\bar\nu}/c$ (since the
antineutrino has very low, if any, mass)

There is no preferred direction for ${\mathbf p}$ and the phase
space (per unit volume) for the electron momentum to lie between
$p$ and $p+dp$ and the antineutrino momentum to lie between $q$ and
$q+dq$ is
  \begin{equation}
    dn_e dn_{\bar\nu} = \frac{4\pi p^2}{h^3}dp\frac{4\pi q^2}{h^3}dq.
  \end{equation}
Our goal is to determine the distribution of the momentum of the
electron.  To this end we integrate $dn_e dn_{\bar\nu}$ over $q$
but subject to the constraint $qc = Q-T$ and find
  \begin{equation}
    dn_e(p) \propto p^2 (Q-T)^2 dp.
  \end{equation}
We transform this distribution to the differential energy distribution
of the electron, $dn_e(E)/dE$,  using the usual rule of transformations
  \begin{eqnarray}
    \frac{dn_e(E)}{dE} &=& \frac{dn_e(p)}{dp} \frac{dp}{dE}\cr
		&\propto& Ep(Q-T)^2.
    \label{eq:dnE}
  \end{eqnarray}			

For the case of a decaying neutron, we use the probability distribution
given by Equation~\ref{eq:dnE} and find that the mean kinetic energy
of the electron is $\langle T\rangle=0.59\, m_e c^2$.  Thus the
heating rate per gram is $N_A\langle T\rangle/t_N = 3.2\times
10^{14}\,$erg\,g$^{-1}$\,s$^{-1}$.

\subsection*{Heating by Decay of Radioactive Nickel}

$^{56}$Ni is produced whenever nuclear reactions are allowed to
proceed to statistical equilibrium. This isotope enjoys the distinction
of being the most packed nucleon (highest binding energy per nucleon).
However, the isotope is unstable and decays as follows:
$^{56}$Ni(6.01\,d)$\rightarrow ^{56}$Co(77\,d)$\rightarrow ^{56}$Fe.
The half lifetimes are given in the parenthesis. The decay proceeds
by electron capture of a K shell electron. Following this capture,
a neutrino is emitted and the $^{56}$Co is left in an excited state,
1.72\,MeV above the ground state. A variety of decays are possible
from this state to the ground state of $^{56}$Co: 1.56\,MeV,
0.812\,MeV, 0.75\,MeV, 0.48\,MeV, 0.27\,MeV and 0.158\,MeV (see
\citealt{a96}, p.\,426).  Cobalt decays by electron capture (80\%)
resulting in a spectrum of $\gamma$-rays of mean energy of 3.5\,MeV
per decay and beta-emission (20\%) with mean energy of 0.14\,MeV
per decay (see \citet{cpk80}).  We set $\langle E_{\rm Co}\rangle
=3.5+0.14=3.64\,$MeV.

Let $N_0=N_{\rm Ni}(t=0)$ be the initial number of Nickel nuclei.
Let $N_{\rm Co}$ be the number of Cobalt nuclei; we have $N_{\rm
Co}(t=0)=0$. The time evolution of Nickel and Cobalt is given by
well known parent-daughter relations:
\begin{eqnarray}
	N_{\rm Ni}(t) &=& N_0\exp(-\lambda_{\rm Ni} t)\ {\rm and} \cr
	N_{\rm Co}(t) &=& N_0\frac{\lambda_{\rm Ni}}{\lambda_{\rm Ni}-
\lambda_{\rm Co}}
		\Bigg(\exp(-\lambda_{\rm Co} t) - \exp(-\lambda_{\rm Ni} t)\Bigg).
\end{eqnarray}
Here, $\lambda_{\rm Ni} = t_{\rm Ni}^{-1}$ and $\lambda_{\rm Co} =
t_{\rm Co}^{-1}$ with $t_{\rm Ni}=6.1/\log(2)$\,d and $t_{\rm
Co}=77/\log(2)$\,d. The heating rates (per gram) are obtained by
letting $N_0=(56 m_H)^{-1}$ and are
\begin{eqnarray}
\varepsilon_{\rm Ni}(t) &=& 
\lambda_{\rm Ni}\Delta E_{\rm Ni} N_{\rm Ni}(t) \cr
&=& 3.9\times 10^{10}\exp(-\lambda_{\rm Ni} t)\,\,{\rm erg\, g^{-1}\, s^{-1}}\cr
\varepsilon_{\rm Co}(t) &=& \lambda_{\rm Co}\langle E_{\rm Co}\rangle
N_{\rm Co}(t)\cr
&=& 7\times 10^{9}\Bigg(\exp(-\lambda_{\rm Co}t)-\exp(-\lambda_{\rm Ni}t)\Bigg)
\,\,{\rm erg\, g^{-1}\, s^{-1}}.
\end{eqnarray}
From an inspection of the heating curve one finds that Cobalt decay
competes with Nickel (assuming full energy absorption) by day 17.
In the main text we note that early time ($\simlt 1$ week) observations
result in the best constraint (or even a discovery). Given this we
can safely ignore the heating provided by the decay of Cobalt.


The next issue is efficiency with which the $\gamma$-rays emitted
during the course of the decay of Nickel can transfer energy to the
surrounding matter.  \citet{cpk80} consider the  deposition of
energy by $\gamma$-rays emanating from decay of Ni$^{56}$ and
Co$^{56}$.  They find that the mass opacity of the $\gamma$-ray
absorption is about 0.029\,cm$^{2}\,$g$^{-1}$, for either Ni$^{56}$
or Co$^{56}$ decay spectrum. They used a Monte Carlo photon transport
code and found that at least for a uniform sphere, $D$, the fraction
of the $\gamma$-ray absorbed by the matter at a given radius depends
only on $\gamma$-ray optical depth (to the surface), $\tau_\gamma$:
   \begin{equation}
	D(\tau_\gamma) = G[1+2G(1-G)(1-0.75G)]
   \end{equation}
where $G=\tau_\gamma/(1.6+\tau_\gamma)$.

I integrated the net deposition fraction for spheres with
center-to-surface optical Thompson optical depth, $\tau_{\rm es}$
ranging from 1 to $10^{17}$.  A polynomial fit yields
\begin{equation}
  \log(\eta(\tau_{\rm es})) = -0.0003z^4 + 0.0108z^3  -0.1486z^2 +  
                               0.8134 z  -1.4360
   \label{eq:eta}
  \end{equation}
where $z=\log(\tau_{\rm es})$; here $\log$ refers to logarithm to the
base 10.

For example, when $\tau_{\rm es}=10$ I find $\eta_\gamma=0.2$ i.e.
80\% of the $\gamma$-rays escape to the surface. Even for $\tau_{\rm
es}=100$ I find $\eta_\gamma=0.7$ i.e. about 30\% escape to the
surface (mainly the outer layers).

\end{document}